\documentstyle[epsf]{article}

\begin{document}
\centerline{\bf Search for fast correlated X-ray and optical variability}
\centerline{\bf in Cir X-1 and XTE J1746-321}
\vspace{2\baselineskip}
\centerline{H.C.\ Spruit}
\centerline{\it Max Planck Institute for Astrophysics, Box 1371, 85741 Garching,
Germany}
\centerline{H.\ Steinle, G.\ Kanbach }
\centerline{\it Max Planck Institute for Extraterrestrial Physics,}
\centerline{\it Box 1312, 85741 Garching, Germany}

\vspace{2\baselineskip}

\leftline{\bf Abstract}
Coordinated observations X-ray+optical observations of two southern X-ray binaries,
the black hole candidate XTE J1746-321 and the neutron star accreter Cir X-1 (a
`microquasar') are reported. With a photon counting optical photometer on the 1.9m
telescope at Sutherland, South Africa and the PCA detector on RXTE, 4h each of
simultaneous data were obtained on XTE J1746 and Cir X-1. Cir X-1 showed no X-ray
variability at the 2\% level,  XTE J1746 was variable at 5-7\% with a 5Hz QPO.
Cross-correlation yielded no correlated signals on either source, to a level of
1\%. A problem with a recently published orbital ephemeris of Cir X-1 is pointed
out.

\section{Introduction}

Though X-rays dominate the energetic output from X-ray binaries, the optical
emission is a potentially powerful diagnostic of the accretion flow, in particular
in combination with the X-ray signal. Reprocessing of X-rays into optical light by
the accretion disk provides information on the geometry (thickness as a function of
distance) and state of ionization of the disk surface. Reprocessing signals of this
kind have been seen in a few neutron star accreters, where the optical `echos'
formed by reprocessing of X-rays from type I X-ray bursts on the neutron star
surface by the disk have been observed (Matsuoka et al. 1984, Turner et al. 1985,
van Paradijs et al. 1990, Kong et al. 2000). 

Optical emission of an entirely different kind has been observed in three black
hole candidates. In  a brief (90s) segment of simultaneous X-ray+optical data of GX
339-4 by Motch et al. (1981,1983) the optical emission was observed to show {\it
dips} some seconds before X-ray peaks. In simultaneous HST+XTE data on GRO 1655-40,
Hynes et al. (1998) found the optical emission to correlate positively with the
X-rays, with a delay of some seconds. The most detailed observations of this kind
were obtained by Kanbach et al. (2001), on the black hole transient XTE J1118+480
(=KV UMa). These observations showed a strong correlation between X-rays and
optical emission, with a cross correlation function showing both a dip preceding X-ray
maximum and a following sharp peak. The shape of the cross correlation function was
found to be highly variable, on time scales as short as 25s (Spruit and Kanbach
2003). One of the most curious properties of this variability is a variation of
time scale: in different segments of data, the cross correlation function tends to
have the same shape, except for expansion/contractraction of the time axis. The
physical interpretation of these observations is still very uncertain.

The short time scales observed in the optical emission in KV UMa  (down to 30ms)
and the large optical luminosity are both suggestive of an origin close to the
black hole. The only radiation process with sufficient emissivity in the optical is
then synchrotron or thermal (cyclo-) synchrotron emission, for which a strong
magnetic field would then be required. Such strong magnetic fields, in turn, are
the ingredient of choice for current models for the production of the relativistic
outflows seen from several black hole transients (Mirabel and Rodr{\'{\i}}guez
1999). These outflows appear to occur in particular when the sources are in X-ray
`hard' states, when the energetic output is dominated by photons around 100keV
(Corbel et al. 2000). 

In contrast with the `soft' state, understood as evidence of the theoretically
predicted of accretion disk structure, the nature of the plasma producing the hard
X-ray emission, and the geometry of the accretion flow in which it occurs is still
unclear. In the few cases where fast correlated X-ray/optical emission has been
found so far, the sources were in such hard states. Correlated X-ray/optical
observations thus provide a potentially powerful new diagnostic on the uncertain
accretion physics close to the hole. 

A difficulty in exploiting this diagnostic is the fact that most black hole
candidates are transients whose outbursts are infrequent and of relatively short
duration. Other difficulties are the tendency for the sources to lie in optically
obscured regions near the galactic center,  and the lack of suitably fast
photometric instrumentation on most modern telescopes.
 
The sources selected for the observations reported here are Cir X-1 and the black
hole transient XTE J1746-321 (=IGR 17464-3213 =H1743-322), which happened to 
be in outburst at the time of the
observations. Cir X-1 is a microquasar (Stewart et al. 1993, Fender et al. 2004) in 
which the accreter is believed to be a neutron star (the other binary of this type
being Sco X-1, Fomalont et al. 2001). Comparison of this object with black
hole candidates would answer the question whether fast optical emission is
connected with the presence of a black hole, or more generically with the presence
of relativistic outflows. Since it is also a persistent X-ray source, it was the
primary target of the observations. 

The second planned target was GX 339-4, the black hole candidate where correlated
X-ray/optical variability was detected for the first time (Motch et al. 1981,1983).
Unlike most black hole candidates, GX 339-4 has frequent outbursts of varying
strength, often several per year. It was active when our observations were planned,
but it turned into an `off' state about 1 month before the observations (as it did
in two previous attempts). However, another X-ray transient and possible black hole
candidate, XTE J1746-321, happened to be active during the observations. Though rather 
obscured, it was still marginally bright enough to attempt optical observations.

\section{Observations}

The optical observations were obtained in the nights of 27 through 31 May 2003 with
the 1.9m Radcliffe telescope of the South African Astronomical Observatory at
Sutherland. Fast photometry was obtained with the OPTIMA photometer, a
photon-counter based on avalanche photo diodes developed at the Max Planck
Institute for Extraterrestrial Physics (Straubmeier et al. 2001). Arrival times of
the photons are recorded with a GPS-based clock. Light was fed to the diode by a
fiber with microlenses at both ends, optimized for the f/18 focus of the telescope.
The effective aperture of the fiber in the focal plane was 4".

On several of the nights the observations were affected by cirrus. Since the
transparency variations occurred on time scales longer than those of interest, the
effect on the results is probably unimportant.

Cir X-1 is a relatively obscured source ($A_{\rm V}\sim 4$, Glass 1994), but still
sufficiently bright  ($I\sim 17.5$) for good photon statistics with the
red-sensitive detector used (to $\sim 950$ nm). Cir X-1 produces X-ray flares
associated with infrared and radio emission around the times of pericenter in its
16.5d excentric orbit\footnote{The new orbital ephemeris given by Parkinson et al.
(2003) appears to be in error. The predicted phases of X-ray flares do not agree
with the ASM light curve. The ephemeris given by Glass (1994) and attributed to
Nicholson, on the other hand, still appears to hold. See also Clarkson et al.
(2004)}. These flares last only a few days, but X-rays continue at a quiescent
level throughout the orbit. Our observations did not cover the pericenter passage.

The optical brightness of XTE J1746-321 (identification and astrometry by Steeghs
et al 2003) is around I=19. The expected optical count rate (around 400 s$^{-1}$)
is below the sky background in the aperture in this case, but still high enough to
extract a signal through cross-correlation with the X-rays, if the source is
sufficiently variable.

Simultaneous X-ray observations were made by the RXTE satellite on each of these
nights for both sources. Total effective simultaneous exposure time was about 4 hrs
each for XTE J1746-321 and Cir X-1.

\begin{figure}
\mbox{}\epsfxsize 0.49\textwidth\epsfbox{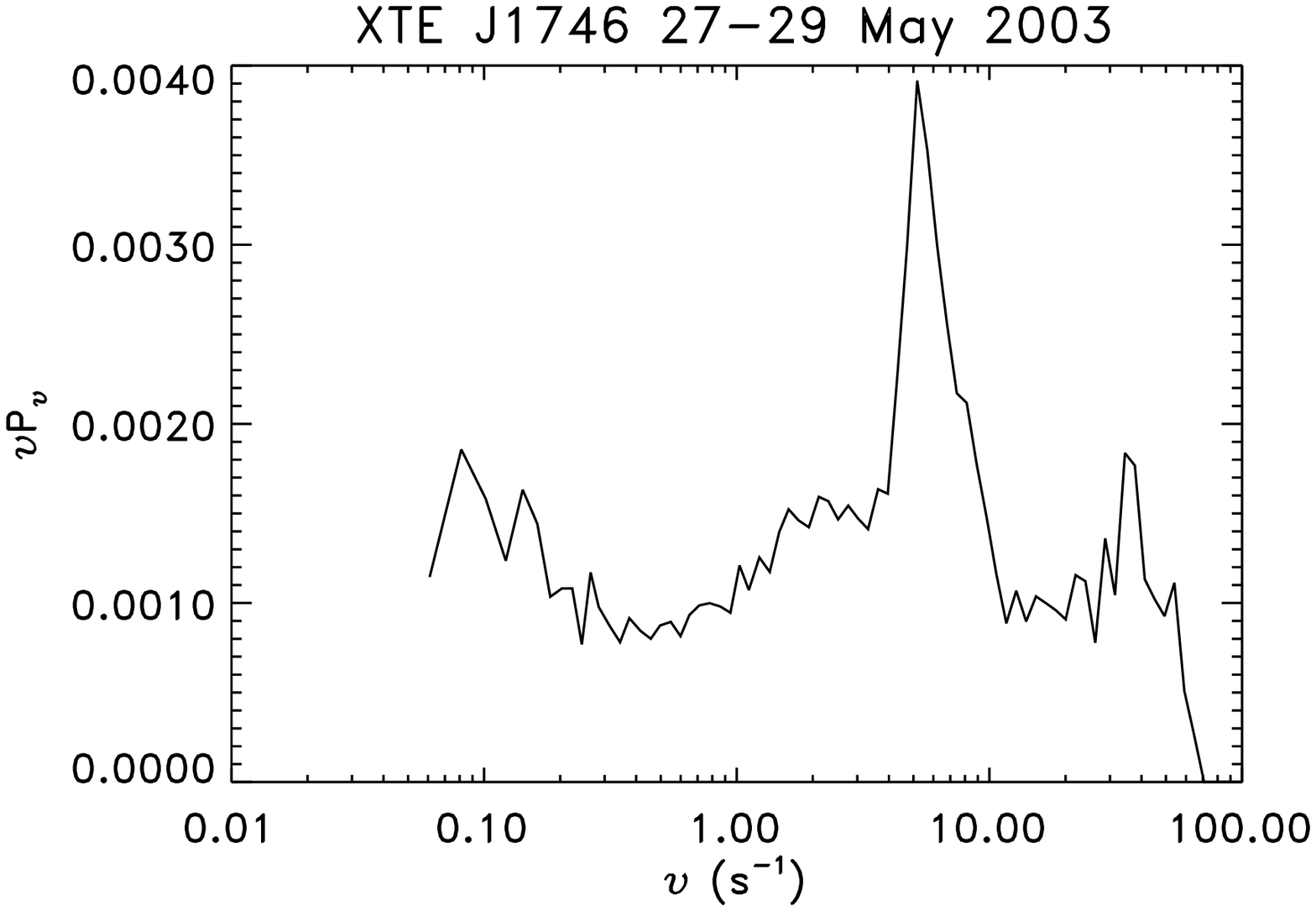}\hfill
\epsfxsize 0.49\textwidth\epsfbox{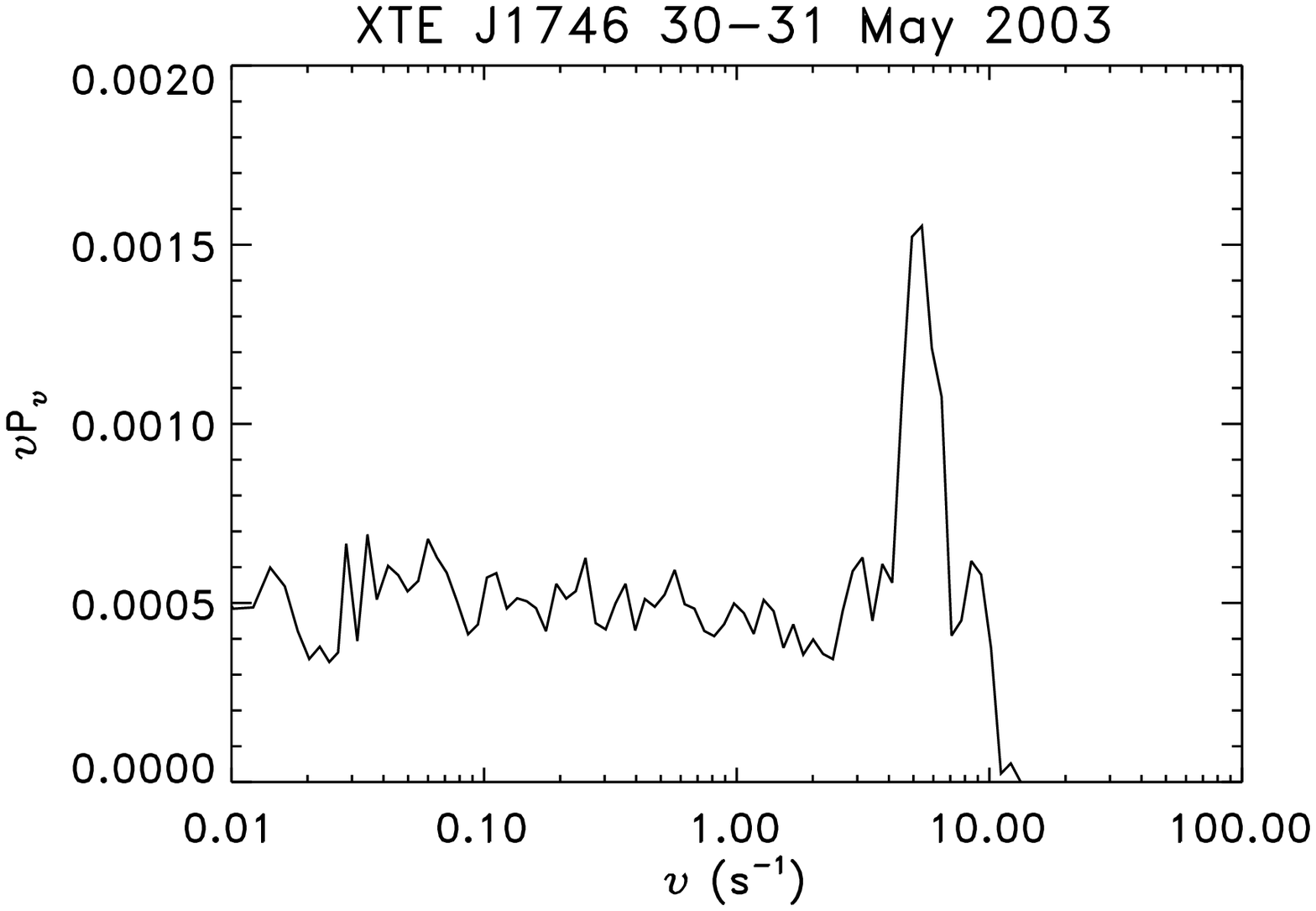}\hfill\mbox{}
\caption{X-ray power spectra ($\nu P_\nu$) of XTE J1746-321 during the first half of the
observations (left panel, bin width 3ms) and the second half (right, bin width 30 ms)}
\end{figure}

\section{Results}

Both the optical and X-ray light curves are dominated by noise. In the case of the
X-rays the noise is just photon statistics, while the noise in the optical signal
was dominated by seeing-related aperture losses. With the long integration times
available, low level intrisic variability can still be extracted out of the noise
by cross-correlation of the X-rays with the optical signal, provided correlated
variability is present. 

The X-ray variability of Cir X-1 happened to be so low as to be barely measurable
in the 4hr integration time; a conservative upper limit is 1.5\% rms (0.01-100 Hz).
In view of this it is not surprising that no significant correlation with the
optical light was detected. The PCA count rate was about 1000 s$^{-1}$ (2 PCUs).

XTE J1746-321 has been mostly in a soft quiet state during its 2003 outburst (e.g. Homan
et al. 2003, but transitions to a hard X-ray state with large variability have been
observed, in particular in September (Grebenev et al. 2003) and October (Tomsick and
Kalemci 2003). Our observations took place during a fairly quiet phase, with modest
variability to correlate the optical signal with (5-8\% rms, 0.1-100Hz). Figure 1 shows 
X-ray power spectra (in $\nu P_\nu$). A quasiperiodic oscillation was present at 5 Hz; 
its amplitude, as well as the overall level of variability was higher during the first
half of the observations (left panel). PCA count rate was about 3000 $s{-1}$ (2 PCUs). 
No significant variability was detected in the optical ($<2$\% rms, 0.1-100 Hz).

No significant cross correlations between X-rays and optical were detected at the
1\% level. This is probably due to the relatively low level of the X-ray
variability, combined with the faintness of XTE J1746-321 in the optical (near the sky 
background).

\section{Discussion}
No significant correlation between the X-rays and optical
signals was detected in either Cir X-1 or XTE J1746-321. In the case of Cir X-1 this 
is probably due to its very low level of X-ray variation at the time of the observations. 
In the case of XTE J1746, a combination of its relatively low X-ray variability and its 
faintness in the optical. Upper limits on correlated X-ray/optical variations are
of the order 1 \% in both Cir X-1 and XTE J1746-321.

These results demonstrate the difficulties of multi-wavelength observing campaigns
at high time resolution. Few observatories are equipped for high time resolution
photometry, also on the southern hemisphere where most of the black hole transients
appear. The use of visitor instruments, as in the observations described here, has
limitations due to the transient nature of the most promising (sufficiently
variable) sources.

\vspace{2\baselineskip}
\leftline{\bf Acknowledgement}
We thank SAAO for the generous allocation of observing time for this project, and
the staff at Sutherland for their hospitality and excellent support of the
observations. We thank Fritz Schrey for hardware development of
the OPTIMA photometer and for assistance with the observations. This research has 
made use of data obtained from the High Energy Astrophysics Science Archive 
Research Center (HEASARC), provided by NASA's Goddard Space Flight Center.

\vspace{2\baselineskip}
\leftline{\bf References}
\parindent=0pt\everypar={\hangindent=2em}

Clarkson,  W.I., Charles, P.A., \& Onyett, N.\ 2004, MNRAS, 348, 458 

Corbel, S., Fender, R.P., Tzioumis, A.K., Nowak, M., McIntyre, V., Durouchoux, P.,
\& Sood, R.\ 2000, A\&A, 359, 251 

Fender, R., Wu, K., Johnston, H., Tzioumis, T., Jonker, P., Spencer, R., \& van der 
Klis, M.\ 2004, Nature, 427, 222 

Fomalont, E.~B., Geldzahler, B.~J., \& Bradshaw, C.~F.\ 2001, ApJ, 558, 
283 

Glass, I.S.\ 1994, MNRAS, 268, 742 

Grebenev, S.A., Lutovinov, A.A., Sunyaev, R.A., 2003, ATEL \#198

Homan, J., Miller, J.M., Wijnands, R., Steeghs, D., et al., 2003, ATEL \# 162

Hynes, R.I., O'Brien, K., Horne, K., Chen, W. \& Haswell, C.A. 1998, 
MNRAS 299, L37

Kanbach, G., Straubmeier, C. Spruit, H.C. \& Belloni, T. 2001 Nature 414, 180

Matsuoka, M.~et al.\ 1984, ApJ, 283, 774 

Mirabel, I.F. \& Rodr{\'{\i}}guez, L.F.\ 1999, ARAA, 37, 409 

Motch, C., Ilovaisky, S.A. \& Chevalier, C. 1981, IAU Circ. 3609

Motch, C., Ricketts, M.J., Page, C.G., Ilovaisky, S.A. \& Chevalier, C. 1983, A\&A
119, 171

Parkinson, P.M.S., et al.\ 2003, ApJ, 595, 333 

Spruit, H.C. \&  Kanbach, G.\ 2002, A\&A, 391, 225 

Steeghs, D., Miller, J.M., Kaplan, D., \& Rupen,M., 2003, IGR/XTE J17464-3213: New
radio position and optical counterpart, ATEL \#146

Stewart, R.T., Caswell, J.L., Haynes, R.F., \& Nelson, G.J.\ 1993, 
MNRAS, 261, 593 

Straubmeier, C., Kanbach, G., \& Schrey, F.\ 2001, Experimental Astronomy, 
11, 157 

Tomsick, J.A., Kalemci, E., 2003, H 1743-322 (= IGR J17464-3213) Transition to the
Hard State, ATEL \# 198.

Turner, M.~J.~L., Breedon, L.~M., Ohashi, T., Courvoisier, T. \& Inoue, H.\ 1985,
SSRv, 40, 249 

van Paradijs, J., van der Klis, M., van Amerongen, S. et al. 1990, A\&A, 234, 181 

\end{document}